%% file: template.tex
\documentclass{article}

\usepackage{arxiv}

\usepackage[utf8]{inputenc} 
\usepackage[T1]{fontenc}    
\usepackage{hyperref}       
\usepackage{url}            
\usepackage{booktabs}       
\usepackage{amsfonts}       
\usepackage{nicefrac}       
\usepackage{microtype}      
\usepackage{lipsum}
\usepackage{graphicx}
\usepackage{amsmath}
\usepackage[square,sort,comma,numbers]{natbib}

\usepackage{array}
\usepackage{caption}
\graphicspath{ {./images/} }

\usepackage{color,colortbl}
\usepackage{todonotes}

\title{SAM \& SAM 2 in 3D Slicer: SegmentWithSAM Extension for Annotating Medical Images}

\author{
 Zafer Yildiz\textsuperscript{1*}, Yuwen Chen\textsuperscript{2}\thanks{Co-first authors} , Maciej A. Mazurowski\textsuperscript{1,2,3,4}\\
 \textsuperscript{1}Department of Radiology, Duke University \\
 \textsuperscript{2}Department of Electrical and Computer Engineering, Duke University \\
 \textsuperscript{3}Department of Biostatistics \& Bioinformatics, Duke University \\
 \textsuperscript{4}Department of Computer Science, Duke University \\
 \texttt{\{zafer.yildiz, yuwen.chen, maciej.mazurowski\}}@duke.edu \\
}

\begin{document}
\maketitle

\input{sec/abstract}
\input{sec/intro}
\input{sec/methods}
\input{sec/results}
\input{sec/conclusion}

\clearpage
\bibliographystyle{splncs04}
\bibliography{references}
\end{document}

%% file: sec/abstract.tex
\begin{abstract}

Creating annotations for 3D medical data is time-consuming and often requires highly specialized expertise. Various tools have been implemented to aid this process. Segment Anything Model 2 (SAM 2) offers a general-purpose prompt-based segmentation algorithm designed to annotate videos. In this paper, we adapt this model to the annotation of 3D medical images and offer our implementation in the form of an extension to the popular annotation software: 3D Slicer. Our extension allows users to place point prompts on 2D slices to generate annotation masks and propagate these annotations across entire volumes in either single-directional or bi-directional manners. Our code is publicly available on \url{https://github.com/mazurowski-lab/SlicerSegmentWithSAM} and can be easily installed directly from the Extension Manager of 3D Slicer as well.

\end{abstract}

%% file: sec/intro.tex
\section{Introduction}
\label{sec:intro}

Medical image segmentation, an essential aspect of medical imaging, involves the accurate delineation of anatomical structures and pathological regions using various imaging modalities such as computed tomography (CT), magnetic resonance imaging (MRI), positron emission tomography (PET), ultrasound \citep{guo2019neutrosophic} and X-ray. Accurate segmentation is fundamental to enhancing diagnostic accuracy, optimizing treatment planning, and ultimately improving patient outcomes \citep{obuchowicz2024clinical,ma2024segment}. Current deep learning methods have demonstrated exceptional capability in automating the segmentation process \citep{rayed2024deep,liu2021review,wang2022medical}. However, training such automatic models still heavily relies on manual expert annotation, which is expensive and time-consuming. To expedite the annotation process, one of the strategies is to leverage the trained models to assist the annotation \citep{tajbakhsh2020embracing,zhang2023efficient}.

Utilizing a finetuned model encounters difficulties due to the wide variety of medical image modalities, often requiring retraining for each modality \citep{zhu2023melo}. Segment Anything Model (SAM) \citep{kirillov2023segment}, as a segmentation foundation model, addresses this issue by offering a zero-shot segmentation solution across different modalities. Additionally, SAM incorporates a prompt-based mode to facilitate the interactive segmentation process. To improve the accuracy and efficiency of annotating medical images, practitioners have finetuned SAM on medical image data \citep{gu2024segmentanybone,shen2024fastsam3d}, and integrated the automatic models into 3D slicer \citep{yildiz2024segmentwithsam,shen2024fastsam,semeraro2023tomosam,liu2023samm}, an open-source software platform designed for the analysis and visualization of medical images \citep{fedorov20123d}.

Medical imaging data is often in 3-dimensional (3D), for example computed tomography (CT) or magnetic resonance imaging (MRI). To efficiently capture the spatial relationships between slices in 3D medical imaging volumetric data and 2D frames in medical videos, practitioners have adapted SAM to handle 3D segmentation tasks, for instance, SAM3D \citep{bui2023sam3d}, SAM-Geo3D \citep{zhang2024samgeod} and SAM-Med3D \citep{wang2023sammed3d}. Recently, SAM 2 was released, enhancing the segmentation process in videos (3D) by incorporating a memory bank to retain information from past predictions \citep{ravi2024sam}. Unlike other 3D models based on SAM, which typically incorporates 3D convolutional block, SAM 2 fundamentally addresses this issue by extending the SAM backbone to 3D. Some studies \citep{dong2024segment} evaluate the capability of SAM 2 in segmenting medical imaging data.


In this paper, we describe a plugin to integrate SAM and SAM 2 into 3D slicer, allowing users to interactively create annotations using hybrid prompt modes in a familiar software environment. Our contributions are summarised below:
\begin{itemize}
    \item We integrate SAM 2 into 3D slicer, which enables propagating annotations in both single and bi-direction starting from any slice in 3D volumes.
    \item We retain 2D promptable segmentation feature we described in \citep{yildiz2024segmentwithsam}, for both SAM and SAM 2.
    \item Our extension involves the usage of all checkpoints provided by SAM and SAM 2.
\end{itemize}

%% file: sec/methods.tex
\section{Methodology}

\begin{figure*}[h]
    \centering
    \includegraphics[width=\linewidth]{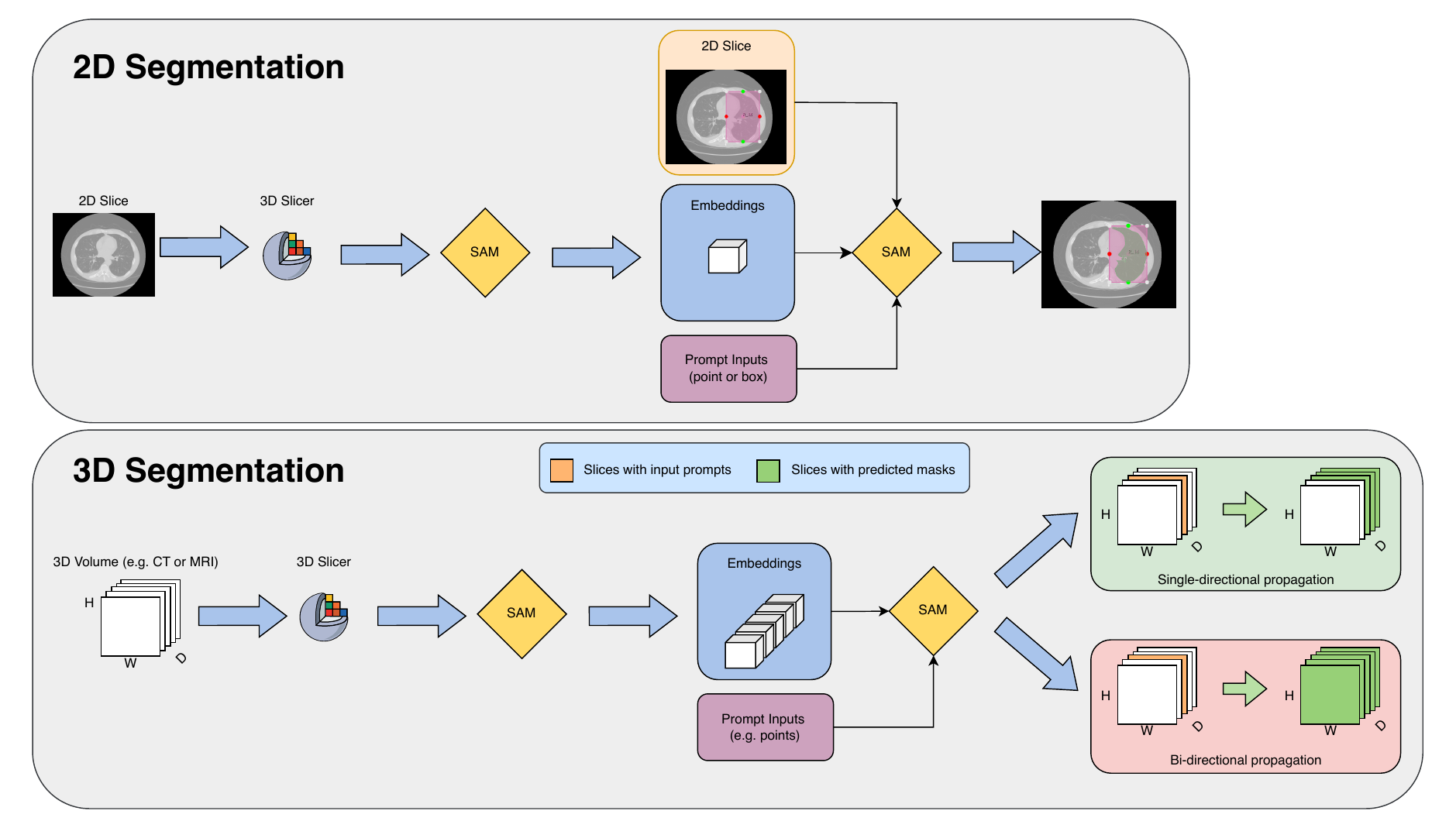}
    \caption{Overview of pipeline for SegmentWithSAM}
    \label{fig:pipeline_segmentwithsam}
\end{figure*}

In this study, we aimed to extend \citep{yildiz2024segmentwithsam}, which aspires to integrate SAM into 3D Slicer for 2D segmentation of medical data, by adding 3D segmentation with SAM 2 model. 3D Slicer's user interface is used as the communication channel between the users and SAM models. The users are expected to open their medical data in 3D Slicer and to give required prompts for any slice they want to segment. The users can either segment one slice by using the 2D image predictor of any SAM or SAM 2 models. SAM 2 also offers video segmentation capability in addition to 2D image segmentation. It enables its users to track single or multiple objects through the frames of a video when proper prompt inputs are given. Thus, this capability makes it a potential tool for segmenting objects of interest in 3D data samples. 

We particularly focused on how we can benefit from SAM and SAM 2's segmentation abilities for segmenting both 2D and 3D medical data in the 3D Slicer user interface. To achieve that we provide 2 options to the users, which are 2D and 3D segmentation:

\subsection{2D Segmentation}

The users are able to segment 2D slices of 3D medical images when they give prompt inputs in 3D Slicer. In this option, all the checkpoints of both SAM (ViT-H, ViT-L, ViT-B) and SAM 2 (Tiny, Small, Base Plus, Large) are available for the users. The users are allowed to select a specific checkpoint before they start segmentation, based on their needs and computational resources. Our extension generates the embedding files for each slice when the users click the start segmentation button for the current slice. This step is only performed once and all the embedding files for each slice are saved so the users do not need to wait every time they want to segment a new slice. After this step, the users are required to place either one or more prompt points or prompt boxes for SAM or SAM 2 to produce the segmentation mask for the current slice. They are allowed to do prompt engineering by adjusting the coordinate and size of prompt inputs to observe the effect on the produced segmentation mask in almost real-time. The users can stop and save the segmentation of the current slice when they get an adequate segmentation mask. In this way, 3D medical images can be segmented by repeating these steps for each slice.

\subsection{3D Segmentation}

To approach 3D medical imaging data such as CT or MRI as a video, we treated each slice as a frame of a video. By this means, the users can segment all the slices of the 3D medical data by using the video predictor of only SAM 2 models after they place prompts on conditional slices (i.e. slices on which prompts are input by users). If they want to segment multiple slices that depend on the conditional slices, they can propagate the segmentation on the conditional slices to the other slices. We provided 2 different propagation modes in our extension, which are propagation through all slices and propagation toward either left or right:

\subsubsection{Propagation through all slices:} In this mode, the users can propagate the produced mask to other slices in a bi-directional way after they place point prompts on any slice they want. If they are not satisfied with the result after the propagation, they can keep adding more prompts on the slices with poor segmentation performance. In this mode, SAM 2 takes new prompt points into account, while it also memorizes the old prompts. This mode might be beneficial for users who want to start segmenting 3D medical data from the middle slices.

\subsubsection{Propagation to left or right:} In this mode, the users can propagate the produced mask either to left or to right. This mode is more useful when the users want to segment a certain number of slices at one end. For this mode, SAM 2 is reset and does not take the previous prompt points into account, unlike the first mode. This mode can be used in combination with the first mode as well. For instance, after the users use the first mode if only a few slices at the right end have poor segmentation, they may prefer to put prompts on the first slice that has a poor segmentation mask and propagate to the right to get a better overall segmentation.

\subsection{Refinement}

Although SAM and SAM 2 provide masks for the segmentation of medical images, there might be errors or noises even after the users give more prompts to SAM and SAM 2. At this point, manual segmentation instruments are still needed to obtain the final segmentation result. Thanks to the strong connection between SAM checkpoints and 3D Slicer, the users can still refine the produced masks by using manual annotation tools of 3D Slicer. As a future work, we also plan to add support for box and mask inputs for SAM 2's video predictor in the upcoming versions of the extension.

%% file: sec/results.tex
\section{Results}
\label{sec:results}

\begin{figure*}[t]
    \centering
    \includegraphics[width=\linewidth]{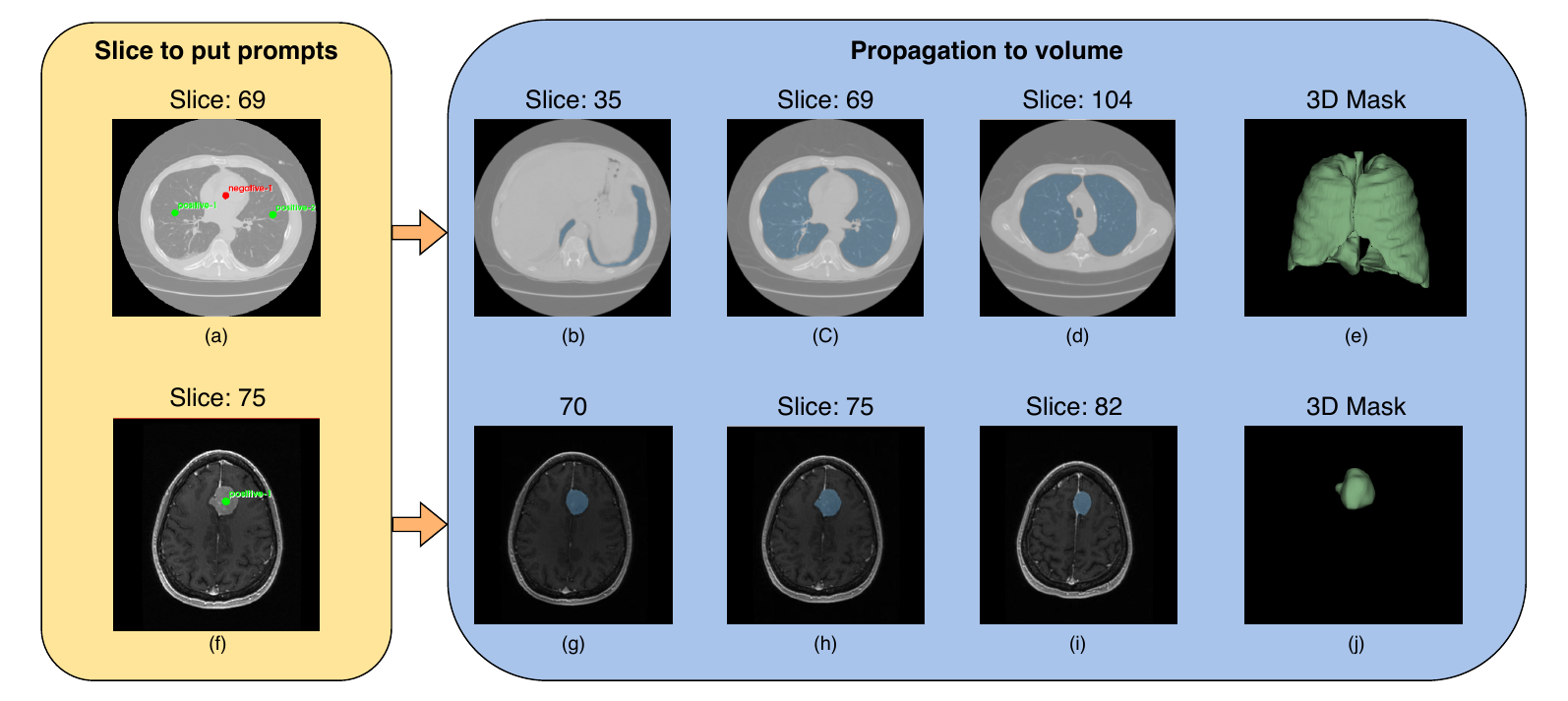}
    \caption{Performance of SegmentWithSAM on different modality samples}
    \label{fig:pipeline_segmentwithsam_result}
\end{figure*}

We evaluate the performance of SAM 2 using publicly available medical data samples across different modalities. For the evaluation, we utilized the sample data from \citep{wiki:xxx,fedorov20123d}. Our evaluation focused on segmenting the liver, where we provide three point prompts (two positives and one negative) placed in the middle slices of the CT scan, as shown in Figure 2(a). For the MRI data, we aim to segment the tumor. We place one point prompt on the slice at the center of the object of interest as depicted in Figure 2(e).

For each sample, after placing the prompts, we generate the segmentation in a bi-directional manner. We present the segmentation results across different slices of the volume: quarter slices (Figure 2(b) and 2(f)), middle slices (Figure 2(c) and 2(g)), third-quarter slices (Figure 2(d) and 2(h)), and the final 3D mask predictions. The segmented masks have shown performance of SegmentWithSAM extension as a valuable tool.

%% file: sec/conclusion.tex
\newpage
\section{Conclusion}
\label{sec:conclusion}

To conclude, we have developed the extension in 3D slicer for SAM 2, facilitating the annotation process on medical imaging data. Different from our previous extension \cite{yildiz2024segmentwithsam} which was based on only SAM, this new extension involves 3D segmentation, a feature enabled by SAM 2. Detailed installation instructions are available on GitHub, and using the extension requires no prior knowledge of coding or deep learning.

%% file: template.bbl
\begin{thebibliography}{10}
\providecommand{\url}[1]{\texttt{#1}}
\providecommand{\urlprefix}{URL }
\providecommand{\doi}[1]{https://doi.org/#1}

\bibitem{bui2023sam3d}
Bui, N.T., Hoang, D.H., Tran, M.T., Le, N.: Sam3d: Segment anything model in volumetric medical images. arXiv preprint arXiv:2309.03493  (2023)

\bibitem{dong2024segment}
Dong, H., Gu, H., Chen, Y., Yang, J., Mazurowski, M.A.: Segment anything model 2: an application to 2d and 3d medical images. arXiv preprint arXiv:2408.00756  (2024)

\bibitem{fedorov20123d}
Fedorov, A., Beichel, R., Kalpathy-Cramer, J., Finet, J., Fillion-Robin, J.C., Pujol, S., Bauer, C., Jennings, D., Fennessy, F., Sonka, M., et~al.: 3d slicer as an image computing platform for the quantitative imaging network. Magnetic resonance imaging  \textbf{30}(9),  1323--1341 (2012)

\bibitem{gu2024segmentanybone}
Gu, H., Colglazier, R., Dong, H., Zhang, J., Chen, Y., Yildiz, Z., Chen, Y., Li, L., Yang, J., Willhite, J., et~al.: Segmentanybone: A universal model that segments any bone at any location on mri. arXiv preprint arXiv:2401.12974  (2024)

\bibitem{guo2019neutrosophic}
Guo, Y., Ashour, A.S.: Neutrosophic sets in dermoscopic medical image segmentation. In: Neutrosophic set in medical image analysis, pp. 229--243. Elsevier (2019)

\bibitem{kirillov2023segment}
Kirillov, A., Mintun, E., Ravi, N., Mao, H., Rolland, C., Gustafson, L., Xiao, T., Whitehead, S., Berg, A.C., Lo, W.Y., et~al.: Segment anything. In: Proceedings of the IEEE/CVF International Conference on Computer Vision. pp. 4015--4026 (2023)

\bibitem{liu2021review}
Liu, X., Song, L., Liu, S., Zhang, Y.: A review of deep-learning-based medical image segmentation methods. Sustainability  \textbf{13}(3), ~1224 (2021)

\bibitem{liu2023samm}
Liu, Y., Zhang, J., She, Z., Kheradmand, A., Armand, M.: Samm (segment any medical model): A 3d slicer integration to sam. arXiv preprint arXiv:2304.05622  (2023)

\bibitem{ma2024segment}
Ma, J., He, Y., Li, F., Han, L., You, C., Wang, B.: Segment anything in medical images. Nature Communications  \textbf{15}(1), ~654 (2024)

\bibitem{obuchowicz2024clinical}
Obuchowicz, R., Strzelecki, M., Pi{\'o}rkowski, A.: Clinical applications of artificial intelligence in medical imaging and image processing—a review. Cancers  \textbf{16}(10), ~1870 (2024)

\bibitem{ravi2024sam}
Ravi, N., Gabeur, V., Hu, Y.T., Hu, R., Ryali, C., Ma, T., Khedr, H., R{\"a}dle, R., Rolland, C., Gustafson, L., et~al.: Sam 2: Segment anything in images and videos. arXiv preprint arXiv:2408.00714  (2024)

\bibitem{rayed2024deep}
Rayed, M.E., Islam, S.S., Niha, S.I., Jim, J.R., Kabir, M.M., Mridha, M.: Deep learning for medical image segmentation: State-of-the-art advancements and challenges. Informatics in Medicine Unlocked p. 101504 (2024)

\bibitem{semeraro2023tomosam}
Semeraro, F., Quintart, A., Izquierdo, S.F., Ferguson, J.C.: Tomosam: a 3d slicer extension using sam for tomography segmentation. arXiv preprint arXiv:2306.08609  (2023)

\bibitem{shen2024fastsam3d}
Shen, Y., Li, J., Shao, X., Romillo, B.I., Jindal, A., Dreizin, D., Unberath, M.: Fastsam3d: An efficient segment anything model for 3d volumetric medical images. arXiv preprint arXiv:2403.09827  (2024)

\bibitem{shen2024fastsam}
Shen, Y., Shao, X., Romillo, B.I., Dreizin, D., Unberath, M.: Fastsam-3dslicer: A 3d-slicer extension for 3d volumetric segment anything model with uncertainty quantification. arXiv preprint arXiv:2407.12658  (2024)

\bibitem{tajbakhsh2020embracing}
Tajbakhsh, N., Jeyaseelan, L., Li, Q., Chiang, J.N., Wu, Z., Ding, X.: Embracing imperfect datasets: A review of deep learning solutions for medical image segmentation. Medical image analysis  \textbf{63},  101693 (2020)

\bibitem{wang2023sammed3d}
Wang, H., Guo, S., Ye, J., Deng, Z., Cheng, J., Li, T., Chen, J., Su, Y., Huang, Z., Shen, Y., Fu, B., Zhang, S., He, J., Qiao, Y.: Sam-med3d (2023), \url{https://arxiv.org/abs/2310.15161}

\bibitem{wang2022medical}
Wang, R., Lei, T., Cui, R., Zhang, B., Meng, H., Nandi, A.K.: Medical image segmentation using deep learning: A survey. IET image processing  \textbf{16}(5),  1243--1267 (2022)

\bibitem{wiki:xxx}
Wiki, S.: Sampledata --- slicer wiki{,} (2019), \url{https://www.slicer.org/w/index.php?title=SampleData&oldid=62556}, [Online; accessed 27-August-2024]

\bibitem{yildiz2024segmentwithsam}
Yildiz, Z., Gu, H., Zhang, J., Yang, J., Mazurowski, M.A.: Segmentwithsam: 3d slicer extension for segment anything model (sam). In: Medical Imaging with Deep Learning (2024)

\bibitem{zhang2024samgeod}
Zhang, J., Yildiz, Z., Gu, H., Dong, H., Mazurowski, M.A.: {SAM}-geo3d: A geometrical method to extend {SAM} to 3d. In: Medical Imaging with Deep Learning (2024), \url{https://openreview.net/forum?id=s7zGrPfsTR}

\bibitem{zhang2023efficient}
Zhang, L., Chen, Z., Zhang, H., Zaman, F.A., Wahle, A., Wu, X., Sonka, M.: Efficient deep-learning-assisted annotation for medical image segmentation. Authorea Preprints  (2023)

\bibitem{zhu2023melo}
Zhu, Y., Shen, Z., Zhao, Z., Wang, S., Wang, X., Zhao, X., Shen, D., Wang, Q.: Melo: Low-rank adaptation is better than fine-tuning for medical image diagnosis. arXiv preprint arXiv:2311.08236  (2023)

\end{thebibliography}
